\documentclass[a4paper,twocolumn,aps,pra]{revtex4}
\usepackage[latin9]{inputenc}
\usepackage{float}
\usepackage{amsmath}
\usepackage{amssymb}
\usepackage{graphicx}
\usepackage{esint}

\makeatletter

\@ifundefined{textcolor}{}
{%
 \definecolor{BLACK}{gray}{0}
 \definecolor{WHITE}{gray}{1}
 \definecolor{RED}{rgb}{1,0,0}
 \definecolor{GREEN}{rgb}{0,1,0}
 \definecolor{BLUE}{rgb}{0,0,1}
 \definecolor{CYAN}{cmyk}{1,0,0,0}
 \definecolor{MAGENTA}{cmyk}{0,1,0,0}
 \definecolor{YELLOW}{cmyk}{0,0,1,0}
}

\usepackage{mathrsfs}\usepackage{subfigure}\usepackage{multirow}\usepackage[title,titletoc,header]{appendix}

\makeatother
\begin{document}

 \draft
 \title{ Double commutator method for  a two  band Bose-Einstein condensate: superfluid density of a flat band superfluid}

  \author{Yi-Cai Zhang}\thanks{ E-mail:zhangyicai123456@163.com}

 \address{School of Physics and Materials Science, Guangzhou University, Guangzhou 510006, China }

\date{\today}

\begin{abstract}

In this work, we propose a double commutator method for a general two-band bosonic superfluid. First, we prove that the sum of the superfluid and normal densities is equal to the weight of the f-sum rule. This weight can be easily determined by analyzing the ground state wave function. Once we have determined the excitation gap of the upper band, we can calculate the normal density by evaluating the average value of a double commutator between the velocity operator and the Hamiltonian. As an application of this method, we investigate the superfluid density of a flat band Bose-Einstein condensate (BEC). Using the Bogoliubov method, we calculate the sound velocity and excitation gap, which allows us to obtain the normal and superfluid densities explicitly. Our findings indicate that the superfluid density is directly proportional to the product of the square of the sound velocity and the compressibility. Furthermore, the existence of a non-vanishing superfluid density depends on the form of the interaction. For example, in the case of U(2) invariant interactions, the superfluid density is zero. Additionally, we have observed that for small interactions, the sound velocity is proportional to the product of the interaction parameter and the square root of the quantum metric of the condensate wave function, which is consistent with the previous literature. However, the superfluid density is directly proportional to the product of the interaction parameter and the quantum metric. The double commutator method indicates that the correction of the excitation gap by interactions is the origin of the non-vanishing superfluid density of a flat band BEC. Up to the linear order of the interaction parameters, all the results for the excitation gap, the normal and superfluid densities in a flat band BEC can also be obtained through a simple perturbation theory. Our work provides another unique perspective on the superfluid behavior of a flat band BEC.

\end{abstract}

\maketitle
\section{Introduction }
A significant number of novel physical phenomena, such as the existence of localized flat band states \cite{Sutherland1986,Vidal1998,Vicencio,Mukherjee}, ferro-magnetism transition \cite{Mielke1999,Zhang2010}, super-Klein tunneling \cite{Shen2010,Urban2011,Fang2016,Ocampo2017}, preformed pairs \cite{Tovmasyan2018}, strange metal \cite{Volovik2019}, and high $T_c$ superconductivity/superfluidity \cite{Peotta2015,Hazra2019,Cao2018,Wuyurong2021,Kopnin2011,Julku2020,Iglovikov2014,Julku2016,Liang2017,Iskin2019,Wu2021,Huhtinen2022,Timo2019,Bernevig}, can appear in a flat band system. Due to the infinitely large density of states of a flat band, a short-ranged potential can result in an infinite number of bound states, including a hydrogen atom-like energy spectrum, where $E_n\propto1/n^2$ for $n=1,2,3,...$ \cite{Zhangyicai2021,Zolotaryuk}. However, a long-range Coulomb potential can completely destroy the flat band \cite{Gorbar2019,Pottelberge2020}. Additionally, it can lead to wave function collapse \cite{Han2019,Zhangyicai20212}, a $1/n$ energy spectrum \cite{Zhangyicai20213}, and even the formation of bound states in a continuous spectrum (BIC) \cite{Zhangyicai20214}.

The superconductivity and superfluidity of fermions with attractive interactions in a flat band lattice system have been extensively investigated \cite{Peotta2015,Hazra2019,Cao2018,Wuyurong2021,Kopnin2011,Julku2020,Iglovikov2014,Julku2016,Liang2017,Iskin2019,Wu2021}. It has been found that the resulting superfluid order parameters are proportional to the strength of the interaction, which is in stark contrast to the exponentially small counterparts found in typical BCS superconductors/superfluids. In an isolated flat band, the superfluid density is directly proportional to the interaction strength when it is weak \cite{Julku2016,Julku2020}. However, in a touching flat band, the superfluid density exhibits a logarithmic singularity as the interaction approaches zero \cite{Iskin2019,Wu2021}. On the other hand, when the interaction is strong, the superfluid density is inversely proportional to the interaction strength, which is related to the effective mass of tightly bound particle pairs \cite{Wu2021}. Additionally, there are gapless phonons that are characterized by total density oscillations, and gapped Leggett modes that correspond to relative density fluctuations between sublattices. Furthermore, it has been discovered that the two-particle spectral functions satisfy an exact sum rule, which is directly related to the filling factor (or particle density) \cite{Wu2021}.

Compared to fermion superfluid, the literature on bosonic superfluid, such as  Bose-Einstein condensate (BEC) in a flat band,  are relatively less \cite{Sebastian2010,youyi, Julku2021,Julku2021B,Zahra,Iskin2022,Iskin2023}.
Unlike typical dispersion bands,  due to the infinity large degeneracy of flat band states, the choice of  condensate momentum needs to minimize the interaction energy \cite{youyi}.
Studies have shown that for weakly interacting Bose-Einstein condensate, the sound velocity, the quantum
depletion and non-vanishing superfluid density are related to the quantum metric of the condensate state \cite{Julku2021,Julku2021B}.
In a flat band BEC, the sound velocity is directly proportional to the strength of the interaction \cite{Julku2021B,Iskin2022}, rather than being proportional to the square root of the interaction parameter as in typical dispersion band BEC.

 In this article, in order to  offer a unique perspective on the superfluid properties of a flat band BEC, based on f-sum rule,  we propose a double commutator method to calculate the superfluid density. Firstly, we present an exact relation between the normal density, superfluid density, and the weights in the f-sum rule. Next, we demonstrate that the normal density can be determined by the average value of a double commutator and the excitation gap. Furthermore, using the Bogoliubov approximation, we calculate the sound velocity and excitation gap. Then, using the f-sum rule and the double commutator method, we obtain the normal density and superfluid density. Differently from quantum metric perspective, our double commutator method  indicate that the correction of the excitation gap by interactions is the origin of the non-vanishing superfluid density in flat band BEC.
  The double commutator method can be also applied to a general two-band BEC.

 The paper is organized as follows. In Sec.\textbf{II}, we give the definitions for the normal density and superfluid density and prove that the sum of the superfluid and normal densities is equal to the weight of the f-sum rule.
  We calculate the normal density with the average value of a double commutator between the velocity operator and Hamiltonian in Sec.\textbf{III}.
  We apply the double commutator method to a 2D flat band BEC in Sec.\textbf{IV}. A summary is given in Sec.\textbf{V}.

\section{f-sum rule, normal density  and superfluid  density }
In this section, we will give the definitions for the superfluid density, normal density and prove that their sum is equal to the weight of the f-sum rule.
\subsection{f-sum rule }
In this section,  we consider a generic many-body Hamiltonian with two-body interaction, i.e.,
 \begin{align}\label{1}
&H=H_0+V_{int}\notag\\
&H_0=\sum_{\alpha,\beta;\textbf{k}}h_{\alpha\beta}(\textbf{k})c^{\dag}_{\alpha \textbf{k}}c_{\beta \textbf{k}},
\end{align}
where $H_0$ is the single-particle Hamiltonian, $\alpha$ is spin (or sublattice) index  and $V_{int}$ is the two-body interaction energy between particles.
$c_{\alpha \textbf{k}}$ is the annihilation operator for particle in the state $|\alpha,\textbf{k}\rangle$.

The density fluctuation operator is given by
   \begin{align}\label{2}
\rho_\textbf{q}= \sum_{\alpha \textbf{k}}c^{\dag}_{\alpha \textbf{k}}c_{\alpha \textbf{k}+\textbf{q}}.
\end{align}
The f-sum rule is given by the following commutator \cite{Pines}, i.e.,
\begin{align}\label{3}
f(\mathbf{q})=[\rho_\textbf{q},[H,\rho_{-\textbf{q}}]].
\end{align}
In addition, we assume that the commutator between the density fluctuation $\rho_q$ and interaction $V_{int}$ vanishes, i.e., $[\rho_\textbf{q},V_{int}]=0$, which is usual situation. Then, we can calculate  the f-sum rule explicitly, i.e.,
\begin{align}\label{6}
 &f(\mathbf{q})=[\rho_\textbf{q},[H,\rho_{-\textbf{q}}]]=[\rho_\textbf{q},[H_0,\rho_{-\textbf{q}}]] \notag\\
 &=\sum_{\alpha \beta \textbf{k}} [ h_{\alpha\beta}(\textbf{k}+\textbf{q})+h_{\alpha\beta}(\textbf{k}-\textbf{q})-2h_{\alpha\beta}(\textbf{k})]  c^{\dag}_{\alpha \textbf{k}}c_{\beta \textbf{k}}.
\end{align}
For free-particle quadratic dispersion Hamiltonian, i.e., $h_0(\mathbf{k})=k^2/(2m)$, f-sum rule becomes
\begin{align}
f(\mathbf{q})=Nq^2/m,
 \end{align}
 where $N=\sum_{\alpha \textbf{k}}c^{\dag}_{\alpha \textbf{k}}c_{\alpha \textbf{k}}$ is total particle number operator and $m$ is single particle's mass.
This implies  that the f-sum rule only depends on the form of the single particle Hamiltonian $H_0$, rather on the interaction $V_{int}$.

Furthermore,  let us take a limit of $\mathbf{q}\rightarrow0$ in Eq.(\ref{6}) and use Taylor expansion near $\mathbf{q}=0$, the f-sum rule is reduced to
\begin{align}
 &f(\mathbf{q})=\sum_{\alpha \beta \textbf{k}} [ h_{\alpha\beta}(\textbf{k}+\textbf{q})+h_{\alpha\beta}(\textbf{k}-\textbf{q})-2h_{\alpha\beta}(\textbf{k})]  c^{\dag}_{\alpha \textbf{k}}c_{\beta \textbf{k}}\notag\\
 &=\sum_{\alpha \beta \textbf{k},ij=x,y,z}q^iq^j \frac{\partial ^2h_{\alpha\beta}(\textbf{k})}{\partial k_i \partial k_j}  c^{\dag}_{\alpha \textbf{k}}c_{\beta \textbf{k}}+O (q^4) \notag\\
 &\simeq\sum_{\alpha \beta \textbf{k},ij=x,y,z} q^iq^j \frac{\partial ^2h_{\alpha\beta}(\textbf{k})}{\partial k_i \partial k_j}  c^{\dag}_{\alpha \textbf{k}}c_{\beta \textbf{k}}\notag\\
 &\equiv\sum_{ij}q^iq^j W_{ij},
\end{align}
where we define the weight of  f-sum rule
 \begin{align}\label{5}
 W_{ij}\equiv\sum_{\alpha \beta \textbf{k}} \frac{\partial ^2h_{\alpha\beta}(\textbf{k})}{\partial k_i \partial k_j}  c^{\dag}_{\alpha \textbf{k}}c_{\beta \textbf{k}}=\lim_{\mathbf{q}\rightarrow0}\frac{1}{2}\frac{\partial^2f(\mathbf{q})}{\partial q_i\partial q_j},
\end{align}
which is a rank two tensor operator.
In the next subsection, we give a proof that the sum of the normal density and superfluid density is equal to the weight of the f-sum rule.

\subsection{ normal density, superfluid density }
 The normal density can be defined by the transverse current-current correlation function \cite{Pines, Baym}. Specifically, at zero temperature, the normal density along the x-direction takes the following form \cite{normaldensity}
 \begin{eqnarray}\label{7}
&&\frac{N\rho_{n,xx}}{m\rho}=\lim_{q_{y,z}\rightarrow0,q_{x}\rightarrow0}\sum_{n\neq0}\frac{2|\langle 0 |j_{\mathbf{q},x}|n\rangle|^2}{\omega_{n0}}
\end{eqnarray}
where $|0\rangle$ and $|n\rangle$ are many-body ground and excited states with eigenenergies $E_0$ and $E_n$. $\omega_{n0}=E_n-E_0$ is excitation energy and $j_{\mathbf{q},x}$ is current fluctuation operator along x-direction, $\rho$ is  mass density, and $m$ is single particle's mass.
At the limit of small $\mathbf{q}$, the current fluctuation operator can be obtained from the continuity equation of particle number, i.e.,
 \begin{eqnarray}\label{71}
&j_{\mathbf{q},x}=\sum_{\alpha\beta \mathbf{k}} \frac{\partial h_{\alpha\beta}}{\partial k_x}c^{\dag}_{\alpha \textbf{k}}c_{\beta \textbf{k}}+O(q).
\end{eqnarray}

For the \emph{transverse} current-current correlation,  we should first take the limit of $q_x\rightarrow 0$ and then $q_{y,z}\rightarrow0$.
The order of taking the limit $\mathbf{q}\rightarrow0$ is crucial. This is because it excludes the contribution of longitudinal excitations in Eq.(\ref{7}), for example, the phonon states
\cite{Pines,normaldensity,Martone2021}.
Usually, in a superfluid system, the lowest collective excitation is gapless phonon with a linear dispersion, i.e., $\omega_{q,-}=cq$.
For a multiple-band system, asides from the phonon excitation, there usually exists upper branch gapped  excitation.
The phonon state does not have contribution to the transverse current-current correlation in Eq.(\ref{7}).
However the upper branch excitations usually result in non-vanishing normal density \cite{normaldensity}.

 Considering the zero contribution of phonon states, we can take  $\mathbf{q}=0$  in Eq.(\ref{7}) without worrying about the order of the $\mathbf{q}\rightarrow0$ limit.
Consequently, the current fluctuation  operator can be replaced by total velocity operator in Eq.(\ref{7}), i.e,
\begin{eqnarray}\label{81}
&\lim_{\mathbf{q}\rightarrow0}j_{\mathbf{q},x}\rightarrow V_x\equiv\sum_{\mathbf{k}\alpha\beta}\frac{\partial h_{\alpha\beta}}{\partial k_x}c^{\dag}_{\alpha \textbf{k}}c_{\beta \textbf{k}}\notag\\
&\equiv \sum_{\mathbf{k}\alpha\beta}v_{\alpha\beta,x}(\mathbf{k})c^{\dag}_{\alpha \textbf{k}}c_{\beta \textbf{k}}
\end{eqnarray}
where we define the group velocity along x-direction $v_{\alpha\beta,x}(\mathbf{k})=\frac{\partial h_{\alpha\beta}(\textbf{k})}{\partial k_x}$ and total velocity operator $V_x$.
So the normal density can be expressed by \cite{normaldensity}
\begin{eqnarray}\label{9}
&&\frac{\rho_{n,xx}}{\rho}=\frac{m}{N}\sum_{n\neq0}\frac{2|\langle 0 |V_x|n\rangle|^2}{\omega_{n0}}.
\end{eqnarray}

The superfluid density can be defined by
employing the phase twist method \cite{Fisher}, i.e.,
 \begin{eqnarray}\label{91}
|\Psi'\rangle=e^{i\theta \sum_k x_k /L_x}|\Psi\rangle
\end{eqnarray}
where $|\Psi\rangle$  is many-body ground state wave function, $L_x$ is the length of the
system and $|\Psi\rangle$ obeys the usual periodic boundary conditions.
The many-body wave function $|\Psi'\rangle$ is then characterized by a
phase twist  at two ends of the system, i.e., $\phi(L_x)-\phi(0)=\theta$.

At zero temperature, the change of ground state energy per particle due to phase twist is related to the superfluid density $\rho_{s,xx}$ of x-direction (also referred to as superfluid
weight or superfluid stiffness in the literature) \cite{Lieb,Fisher}
\begin{eqnarray}
 &\frac{E^{'}_0}{N}=\frac{E_0}{N}+\frac{\rho_{s,xx}}{2m\rho}(\frac{\theta}{L_x})^2+O[(\frac{\theta}{L_x})^4]
\end{eqnarray}
where $E'_0$ and $E_0$ are the ground state energies in the presence
and in the absence of the twist constraint, respectively.
When $\theta/L_x$  is small, the superfluid density is
\begin{eqnarray}\label{12}
 \frac{\rho_{s,xx}}{m\rho}=\frac{2(E^{'}_0-E_0)}{N(\frac{\theta}{L_x})^2} \equiv\frac{2\Delta E_0}{N(\frac{\theta}{L_x})^2}
\end{eqnarray}
where  $\Delta E_0$ is the difference between  $E'_0$ and $E_0$.

According to (\ref{91}), the  total momentum operator $P_x$ acts on $|\Psi\rangle$
as $P_x |\Psi\rangle= e^{i\theta \sum_k x_k /L_x}(P_x + N \theta/Lx ) |\Psi\rangle$ and
consequently the calculation of $E'_0$ corresponds to minimizing
the energy with respect to $|\Psi\rangle$ with a modified Hamiltonian:
\begin{eqnarray}
 \langle \Psi' | H(p_x) | \Psi' \rangle=\langle \Psi | H(p_x+ \theta/L_x) | \Psi \rangle.
\end{eqnarray}
This means the momentum should become
\begin{eqnarray}\label{14}
 & p_x\rightarrow p^{'}_{x}\equiv p_x+\theta/L_x,
\end{eqnarray}
and the Hamiltonian is changed as
\begin{eqnarray}
 & H' =H(p_x+ \theta/L_x) =H(p_x)+\Delta H.
\end{eqnarray}

Due to the gauge invariance of interaction $V_{int}$ under the phase twist, by Eq.(\ref{1}), the perturbation is
\begin{align}
&\Delta H=\Delta H_0=\frac{\theta}{L_x}\sum_{\alpha,\beta;\textbf{k}}\frac{\partial h_{\alpha\beta}(\textbf{k})}{\partial k_x} c^{\dag}_{\alpha \textbf{k}}c_{\beta \textbf{k}}\notag\\
&+\frac{1}{2}(\frac{\theta}{L_x})^2\sum_{\alpha,\beta;\textbf{k}}\frac{\partial^2 h_{\alpha\beta}(\textbf{k})}{\partial^2 k_x}c^{\dag}_{\alpha \textbf{k}}c_{\beta \textbf{k}}\notag\\
&=\frac{\theta}{L_x} V_x+\frac{1}{2}(\frac{\theta}{L_x})^2 W_{xx}.
\end{align}

The energy's change $\Delta E_0$ can be easily calculated with the second order perturbation theory, i.e.,
\begin{eqnarray}\label{17}
 &&\Delta E_0=\langle 0|\Delta H|0\rangle+\sum_{n\neq0}\frac{|\langle 0|\Delta H|n\rangle|^2}{E_0-E_n}\notag\\
 &&=\frac{1}{2}(\frac{\theta}{L_x})^2 [\overline{W}_{xx}+2\sum_{n\neq0}\frac{|\langle 0|V_x|n\rangle|^2}{E_0-E_n}].
\end{eqnarray}
In the above equation, we assume there is no net particle flow in ground state, i.e., $\langle 0|V_x|0\rangle=\langle \Psi|V_x|\Psi\rangle=0$.

 By Eq.(\ref{9}), we see that the second term in Eq.(\ref{17}) is proportional the normal density, i.e.,
 \begin{eqnarray}\label{18}
&2\sum_{n\neq0}\frac{|\langle 0|V_x|n\rangle|^2}{E_0-E_n}]=-2\sum_{n\neq0}\frac{|\langle 0|V_x|n\rangle|^2}{E_n-E_0}]\notag\\
&=-2\sum_{n\neq0}\frac{|\langle 0|V_x|n\rangle|^2}{\omega_{n0}}]=-N \rho_{n,xx}/(m\rho).
\end{eqnarray}

By Eqs.(\ref{5}), (\ref{12}), (\ref{17}) and (\ref{18}), we find the sum of $\rho_n$ and $\rho_s$ is proportional to the weight of the f-sum rule, i.e.,
\begin{eqnarray}\label{A2}
\frac{\rho_{s,xx}}{\rho}+\frac{\rho_{n,xx}}{\rho}=\frac{m\overline{W}_{xx}}{N}=\lim_{\mathbf{q}\rightarrow 0}\frac{m\overline{f(\mathbf{q}=q_x\mathbf{e}_x)}}{Nq^{2}_{x}}.
\end{eqnarray}
where we introduce the unit vector of x-direction $\mathbf{e}_x$ .
So if  we know the weight of f-sum rule $\overline{W}_{xx}$ and $\rho_{n,xx}$, by Eq.(\ref{A2}), then we can calculate the superfluid density $\rho_{s,xx}$.
The relation Eq.(\ref{A2}) of normal density, superfluid density and f-sum rule also holds as a tensor identity, i.e.,
\begin{eqnarray}\label{f}
\frac{\rho_{s,ij}}{\rho}+\frac{\rho_{n,ij}}{\rho}=\frac{m\overline{W}_{ij}}{N}=\lim_{\mathbf{q}\rightarrow0}\frac{m}{2N}\frac{\partial^2\overline{f(\mathbf{q})}}{\partial q_i\partial q_j}.
\end{eqnarray}
where $i,j=x,y,z$, the average value and weight of f-sum rule in ground state are
\begin{eqnarray}
&&\overline{f(\mathbf{q})}\equiv\langle 0|f(\mathbf{q})|0\rangle,\notag\\
   &&\overline{W}_{ij}=\sum_{\alpha \beta \textbf{k}} \frac{\partial ^2h_{\alpha\beta}(\textbf{k})}{\partial k_i \partial k_j} \langle0| c^{\dag}_{\alpha \textbf{k}}c_{\beta \textbf{k}}|0\rangle,
\end{eqnarray}
and similar to Eq.(\ref{9}), $\frac{\rho_{n,ij}}{\rho}$ is
\begin{eqnarray}\label{901}
&&\frac{\rho_{n,ij}}{\rho}=\frac{\rho_{n,ji}}{\rho}\notag\\
&&=\frac{m}{N}\sum_{n\neq0}\frac{\langle 0 |V_i|n\rangle\langle n|V_j|0\rangle+\langle 0 |V_j|n\rangle\langle n|V_i|0\rangle}{\omega_{n0}}.
\end{eqnarray}
where $V_i$ is $i-$th spatial direction total velocity operator
\begin{eqnarray}
V_i\equiv\sum_{\mathbf{k}\alpha\beta}\frac{\partial h_{\alpha\beta}}{\partial k_i}c^{\dag}_{\alpha \textbf{k}}c_{\beta \textbf{k}}.
\end{eqnarray}
The diagonal elements of the superfluid density \((\rho_{s,ii})\) and the normal density \((\rho_{n,ii})\) are typically non-negative. According to Eq.(\ref{f}), if the weight of the f-sum rule \(\overline{W}_{ii}\) is equal to zero, then both the superfluid density and the normal density will vanish. Since \(\rho_{n,ii} \geq 0\) (see Eq.(\ref{9})), this implies that the superfluid density \(\rho_{s,ii}\) has an upper bound given by:
 \begin{eqnarray}\label{bound}
\rho_{s,ii}\leq \frac{m\overline{W}_{ii}}{N}=\frac{m}{N}\sum_{\alpha \beta \textbf{k}} \frac{\partial ^2h_{\alpha\beta}(\textbf{k})}{\partial k_i \partial k_i} \langle0| c^{\dag}_{\alpha \textbf{k}}c_{\beta \textbf{k}}|0\rangle,
\end{eqnarray}
which is consistent with the findings of Ref.\cite{Hazra2019}.
Taking into account the stability of the superfluid system, it is important to note that, as a symmetric matrix, all eigenvalues of \(\rho_s\) (or \(\rho_n\)) should be non-negative.

When the total velocity operator $V_i$ and the Hamiltonian $H$ commute, meaning that $[V_i, H] = 0$, we can always select their common eigenstates $|0\rangle$ and $|n\rangle$. In this case, the total velocity is a good quantum number, and the matrix element $\langle 0 | V_i | n \rangle $ is zero. As a result, the normal density given by Eq. (\ref{901}) would also vanish. Therefore, according to Eq. (\ref{f}), the superfluid density corresponds to the weight of the f-sum rule, and the inequality described in Eq. (\ref{bound}) becomes an equality.
However, in a general multi-band system where $[V_i, H] \neq 0$, the normal density is typically not zero.

\section{double commutator  method for normal density}
In this section, we propose a double commutator  method to calculate the normal density $\rho_{n,xx}$ \cite{normaldensity}. First of all, let us calculate the average value of the following double commutator
\begin{eqnarray}\label{201}
&&\langle 0|[V_i,[H,V_j]]|0\rangle\notag\\
&&=\sum_{n\neq0}\omega_{n0}[\langle0|V_i|n\rangle\langle n|V_j|0\rangle+\langle0|V_j|n\rangle\langle n|V_i|0\rangle].
\end{eqnarray}
Furthermore, if there is only one upper branch excitation gap $\omega_{n0}=\Delta$ and there is only one dominant term in the summation of Eqs.(\ref{901}) and (\ref{201}), then
comparing Eq.(\ref{201}) with Eq.(\ref{901}), we find that  the normal density can be given approximately by the average value of double commutator and excitation gap \cite{normaldensity}, i.e.,
\begin{eqnarray}\label{21}
\frac{\rho_{n,ij}}{\rho}=\frac{\rho_{n,ji}}{\rho}\simeq\frac{m\langle 0|[V_i,[H,V_j]]|0\rangle}{N\Delta^2}.
\end{eqnarray}
This method, Eq.(\ref{21}), has been successfully applied to two band spin orbit coupled BEC \cite{normaldensity}.

As long as we know the average value of the double commutator and excitation gap $\Delta$, we can obtain the normal density and then by Eq.(\ref{f}), superfluid density.
Since this method does not require knowledge of the excited state wave function $|n\rangle$, it is typically easier to calculate the average value of the double commutator and the excitation gap compared to directly calculating the current-current correlation function [Eq.(\ref{7})].
Such a method can be applied to a general two-band bosonic  superfluid, for example, spin-orbit coupled BEC \cite{normaldensity}, a flat band BEC (see next section), and so on  \cite{Iskin2023}.

\section{Application: a 2D flat band BEC}
In this section, we consider a 2D two-component (band) Bose-Einstein condensate  described by a flat band Hamiltonian
\begin{eqnarray}\label{eqn1}
     &&H=H_0+V_{int} ,\notag\\
     &&H_0=\int d^3\mathbf{r}\psi^+[\frac{p_{x}^2+p_{y}^{2}}{2m^*}+\frac{p_xp_y\sigma_z}{m^*}+\frac{p_{x}^2-p_{y}^2}{2m^*}\sigma_x]\psi\notag\\
     &&V_{int}=\frac{1}{2}\int d^3\mathbf{r}[g\psi_{1}^\dag(\mathbf{r})\psi_{1}^\dag(\mathbf{r})\psi_{1}(\mathbf{r})\psi_{1}(\mathbf{r})\notag\\
     &&+2g'\psi_{1}^\dag(\mathbf{r})\psi_{2}^\dag(\mathbf{r})\psi_{2}(\mathbf{r})\psi_{1}(\mathbf{r})\notag\\
     &&+g\psi_{2}^\dag(\mathbf{r})\psi_{2}^\dag(\mathbf{r})\psi_{2}(\mathbf{r})\psi_{2}(\mathbf{}r)],
\end{eqnarray}
where $H_0$ and $V_{int}$ are single-particle Hamiltonian and interaction energy between particles, respectively. $\psi_{1(2)}$ are two component bosonic field operators, $\sigma_{z(x)}$ act on spin (or sublattice) space, $m^*$ is effective mass,  $g(g')$ is intra-species (inter-species) interaction strengths between particles.
  $\psi^\dag=[\psi^\dag_1,\psi^\dag_2]$ is field operator in two component spinor form. When  $g=g'$,   the interaction $V_{int}$  is U(2) rotational transformation  invariant in spin space  \cite{zhengwei}. In the whole paper, we assume $g>0$ and $g-g'\geq0$. let us define two interaction parameters:
  \begin{align}
G_1=n(g + g')/2,\ \ \ \ G_2 = n(g-g'),
\end{align}
 where $n=N/V$ is the average particle density, which would be extensively used in the following text.
The free particle part of this model $H_0$ can be viewed as a continuous version of two band Mielke checkerboard lattice \cite{Iskin2019} or singular quadratic touching flat band checkerboard-I/II lattice model  \cite{Rhim}.

 With single-particle Hamiltonian
  \begin{eqnarray}\label{25}
h_0(\mathbf{p})=\frac{p_{x}^{2}+p_{y}^2}{2m^*}+\frac{p_xp_y\sigma_z}{m^*}+\frac{p_{x}^2-p_{y}^2}{2m^*}\sigma_x,
\end{eqnarray}
 the two band  single particle wave functions and eigenenergies are given by
\begin{align}\label{277}
&\psi_-(\mathbf{p})=\left(                 
 \begin{array}{c}   
   \frac{p_x-p_y   }{\sqrt{2(p^{2}_{x}+p^{2}_{y})}} \\  
       \frac{-(p_x+p_y)   }{\sqrt{2(p^{2}_{x}+p^{2}_{y})}}\\    
\end{array}\right)e^{i\mathbf{p}\cdot \mathbf{r}},\notag\\
&\psi_+(\mathbf{p})=\left(                 
 \begin{array}{c}   
   \frac{p_x+p_y   }{\sqrt{2(p^{2}_{x}+p^{2}_{y})}} \\  
       \frac{p_x-p_y   }{\sqrt{2(p^{2}_{x}+p^{2}_{y})}}\\    
\end{array}\right)e^{i\mathbf{p}\cdot \mathbf{r}},\notag\\
&E_{-}(\mathbf{p})=0,\ \ \ \ E_{+}(\mathbf{p})=\frac{p_{x}^{2}+p_{y}^{2}}{m^*},
\end{align}
where $-(+)$ stand for lower and upper band, respectively. It is found that the lower band is completely flat, i.e., $E_{-}(\mathbf{p})\equiv0$ for an arbitrary $\mathbf{p}=[p_x,p_y]$.
At zero temperature, the condensation occurs at the lower band.
  Due to the degeneracy of lower flat band, the condensate momentum needs to be determined by considering the minimization of interaction energy \cite{youyi}.
  By  minimizing the interaction energy $V_{int}$, we find that the lowest mean field energy  can be obtained at momentum $\mathbf{p}_0=[k_0,0]  $ or $\mathbf{p}_0=[0,k_0]$ where $k_0$ is an arbitrary non-zero constant.

  In the following, we will assume that Bose-Einstein condensation occurs at a single momentum $\mathbf{p}_0=[k_0,0]$ and occupies the lower flat band. At the condensate momentum $\mathbf{p}_0=[k_0,0]$, there exists a single-particle excitation gap $\Delta_0$, given by: \begin{align}\label{d} \Delta_0=E_+(\mathbf{p}_0)-E_-(\mathbf{p}_0)=\frac{k^{2}_{0}}{m^*}. \end{align}

  The order parameter, also known as the condensate wave function, is plane-wave state, i.e.,
\begin{align}\label{2721}
\phi=\left(                 
 \begin{array}{c}   
   \phi_1    \\  
   \phi_2       
\end{array}\right)e^{ik_0 x} =\sqrt{\frac{N_0}{2V}}\left(                 
 \begin{array}{cccc}   
   1    \\  
   -1  \\  
\end{array}\right)e^{ik_0 x},
\end{align}
 where $N_0$ is particle's number in condensate, and for weakling interacting bosonic gas, we expect that $N_0\approx N$.
 The mean-field ground state is
\begin{align}\label{270}
|0\rangle=(\sqrt{\frac{1}{2}})^N\prod _{k=1}^{N}\left(                 
 \begin{array}{cccc}   
   1    \\  
   -1  \\  
\end{array}\right)_ke^{ik_0 x_k},
\end{align}
where $N$ is total particle number and $k$ is the index for $k-th$ particle.
We have observed that the ground state exhibits degeneracy in the x-direction of momentum space. This degeneracy cannot be eliminated within the mean-field framework. To address this issue, one may need to consider the contribution of quantum fluctuations to the ground state energy. However, such a calculation is complex and falls outside the scope of our current work. Therefore, our subsequent discussions are based on the assumption that the mean-field ground state wave function, as shown in Eq.(\ref{270}), exists.

 By Eq.(\ref{270}), we see that the ground state wave function is an eigenstate of the spin operator $\sigma_x$ with eigenvalues $-1$.
With condensate wave function Eq.(\ref{270}), we find that the average values of some physical quantities are
\begin{align}\label{272}
& \ \ \langle 0|p_x|0\rangle =k_0,\ \ \langle 0|p_y|0\rangle =0,\notag\\
&\langle0|\sigma_y|0\rangle=\langle0|\sigma_z|0\rangle=0, \ \ \langle0|\sigma_x|0\rangle=-1,
\end{align}
which would be used in the following discussions.

\subsection{f-sum rule  }
Using Eqs.(\ref{2}), (\ref{3}) and  (\ref{25}), after directly calculating, we find that the f-sum rule takes the following form
\begin{eqnarray}
f(\mathbf{q})=\frac{q^2N}{m^*}+\frac{2q_xq_y\Sigma_z}{m^*} +\frac{(q_{x}^{2}-q_{y}^{2})\Sigma_x}{m^*},
\end{eqnarray}
where $q=\sqrt{q_{x}^2+q_{y}^{2}}$, the total spin operator $\Sigma_z\equiv \sum_k \sigma_{k,z}$, $\Sigma_x\equiv\sum_k \sigma_{k,x}$, and  $\sum_{k}$ denotes summation over all the particles.
By Eq.(\ref{272}), the average value of f-sum rule along x-direction is
\begin{eqnarray}\label{271}
\overline{f(\mathbf{q}=q_x\mathbf{e}_x)}=\frac{q^{2}_{x}N}{m^*}+\frac{q_{x}^{2}\overline{\Sigma}_x}{m^*}=0.
\end{eqnarray}
By Eq.(\ref{f}), we find that the sum of normal and superfluid density vanishes, i.e,
\begin{eqnarray}\label{27}
\frac{\rho_{s,xx}}{\rho}+\frac{\rho_{n,xx}}{\rho}=0.
\end{eqnarray}

For y-direction, we have
\begin{eqnarray}\label{281}
\overline{f(\mathbf{q}=q_y\mathbf{e}_y)}=\frac{q^{2}_y N}{m^*}-\frac{q_{y}^{2}\overline{\Sigma}_x}{m^*}=\frac{2Nq_{y}^{2}}{m^*}
\end{eqnarray}
where we introduce unit vector of y-direction $\mathbf{e}_y$ and
\begin{eqnarray}\label{28}
\frac{\rho_{s,yy}}{\rho}+\frac{\rho_{n,yy}}{\rho}=\frac{2m}{m^*}.
\end{eqnarray}
Similarly, we find
\begin{eqnarray}
\frac{\rho_{s,xy}}{\rho}+\frac{\rho_{n,xy}}{\rho}=0,\ \ \ \ \frac{\rho_{s,yx}}{\rho}+\frac{\rho_{n,yx}}{\rho}=0.
\end{eqnarray}

\subsection{Sound velocity and excitation gap }
In the momentum space, the field operator can be written as  $\psi_{1(2)}=\frac{1}{\sqrt{V}}\sum_\mathbf{k} c_{1(2)\mathbf{k}}e^{i\mathbf{k}\cdot \mathbf{r}}$.
 When Bose-Einstein condensation takes place,   the field operator takes following form
 \begin{align}
  \psi_{1(2)}=\phi_{1(2)}+\delta\psi_{1(2)},
\end{align}
 where order parameter $\phi_{1(2)}=\langle c_{1(2)\mathbf{p}_0}\rangle$ is ordinary  number and $\delta\psi_{1(2)}$ is fluctuation around order parameter. Taking into account the fluctuation $\delta\psi_{1(2)}$  within Bogliubov framework, one can obtain Bogoliubov equation \cite{zhengwei,Martone}
\begin{align}
K_{\mathbf{q}}
\left(                 
 \begin{array}{cccc}   
   u \\  
   v \\  
\end{array}\right)=\omega\left(                 
 \begin{array}{cccc}   
   u \\  
   v \\  
\end{array}\right),
\end{align}
where the particle amplitude ($u$) and hole  amplitude ($v$) are $2\times1$ column vectors  and $K_{\mathbf{q}}$ is a $4\times4$ matrix.

The Bogliubov matrix takes the following form
\begin{align}
K_{\mathbf{q}}\!\!=\!\!\left(  \!\!               
 \begin{array}{cccc}   
   h_0(\mathbf{p}_0+\mathbf{q})-\mu+\Sigma_N   \!\!&\!\! \Sigma_A \\  
   -\Sigma_A \!\!&\!\! -(h_0(\mathbf{p}_0-\mathbf{q})-\mu+\Sigma_N )\\  
\end{array}\!\!\right)_{4\times4},
\end{align}
and the chemical potential matrix
\begin{align}
\mu\!\!=\!\!\left( \!\!\!                
 \begin{array}{cccc}   
   gn_1+g'n_2   &0 \\  
   0 & gn_2+g'n_1  \\  
\end{array}\!\!\!\right),
\end{align}
where  $n_{1(2)}$ is the particle number density of first (second) component which satisfy $n_1=n_2=n/2$ and total density $n=n_1+n_2$.
The normal self-energy
\begin{align}
\Sigma_N\!\!=\!\!\left( \!\!\!                
 \begin{array}{cccc}   
   2gn_1+g'n_2   &-g'\sqrt{n_1n_2} \\  
   -g'\sqrt{n_1n_2} & 2gn_2+g'n_1 \\  
\end{array}\!\!\!\right),
\end{align}
and the anomalous self-energy
\begin{align}
\Sigma_A=\left(                 
 \begin{array}{cccc}   
   gn_1   &-g'\sqrt{n_1n_2} \\  
   -g'n\sqrt{n_1n_2} & gn_2 \\  
\end{array}\right).
\end{align}
It is important to note that a generalized Hugenholtz-Pines relation (as a matrix identity) holds \cite{Hugenholtz1959,Kriszti,yicai2018Josephson,Watabe}, i.e.,
\begin{align}
\mu=\Sigma_N-\Sigma_A.
\end{align}
which  guarantees that the low energy excitation of a superfluid is gapless (phonon).

Due to the presence of two components, there exist two branch excitation spectra. One is gapless phonon  (denoting $-$), the other is gapped up-branch excitation (denoting $+$).
The particle operator ($a_{1(2)\mathbf{p}_0\pm \mathbf{q}}$) in momentum space can be expressed by the quasi-particle operator ($b_{q \pm}$)
\begin{align}
\left(                 
 \begin{array}{c}   
   a_{1\mathbf{p}_0+\mathbf{q}}     \\  
   a_{2\mathbf{p}_0+\mathbf{q}}       \\  
   a^{+}_{1\mathbf{p}_0-\mathbf{q}}  \\
   a^{+}_{2\mathbf{p}_0-\mathbf{q}}
\end{array}\right) =U\left(                 
 \begin{array}{c}   
   b_{\mathbf{q}+}     \\  
   b_{\mathbf{q}-}       \\  
   b^{+}_{-\mathbf{q}+}  \\
   b^{+}_{-\mathbf{q}-}
\end{array}\right),
\end{align}
where the Bogoliubov transformation matrix $U$ is made of the the eigenvectors of the above Bogliubov matrix $K_q$,
\begin{align}
U=\left(                 
 \begin{array}{cccc}   
   u_{1\mathbf{q}+}    &    u_{1\mathbf{q}-}    &  v_{1(-\mathbf{q})+} &v_{1(-\mathbf{q})-} \\  
   u_{2\mathbf{q}+}  &    u_{2\mathbf{q}-}    &  v_{2(-\mathbf{q})+}     & v_{2(-\mathbf{q})-} \\  
   v_{1\mathbf{q}+}     &   v_{1\mathbf{q}-}  &   u_{1(-\mathbf{q})+} &u_{1(-\mathbf{q})-} \\
   v_{2\mathbf{q}+}   &   v_{1\mathbf{q}-}     & u_{2 (-\mathbf{q})+}    &  u_{2(-\mathbf{q})-}
\end{array}\right).
\end{align}

Under the transformation $U$, the Bogloiubov matrix takes a diagonal form
\begin{align}
U^{-1}K_{\mathbf{q}}U=\left(                 
 \begin{array}{cccc}   
   \omega_+(\mathbf{q})    &    0    &  0 &0 \\  
   0  &     \omega_-(\mathbf{q})    &  0     & 0 \\  
   0     &   0  &    -\omega_+(-\mathbf{q}) &0 \\
   0   &   0     &0     &  -\omega_-(-\mathbf{q})
\end{array}\right).
\end{align}
 After diagonalizing the matrix $K_{\mathbf{q}}$, we get the gapless phonon and excitation gap, i.e.,
 \begin{align}\label{41}
&&\omega_-(\mathbf{q})=c(\mathbf{\hat{q}})q=\sqrt{\frac{2G_1G_2sin^{2}(\theta)}{m^*(G_2+\Delta_0)}}q,\notag\\
&&\Delta\equiv \omega_+(\mathbf{q}\rightarrow 0)=\sqrt{\Delta_0(G_2+\Delta_0)}.
\end{align}
where $c(\mathbf{\hat{q}})$ is sound velocity  and
\begin{align}\label{322}
\mathbf{\hat{q}}=\mathbf{q}/q=[q_x,q_y]/q=[cos(\theta),sin(\theta)]
\end{align}
 is the unit vector along direction of $\mathbf{q}$.
When $\theta=0$,  then $\mathbf{q}=[q,0]$ and $c=0$.
This implies the sound velocity along x-direction is zero, which reflect the degeneracy of ground states along x-direction of momentum space.
 Comparing Eq.(\ref{41}) with Eq.(\ref{d}),  we see that in the presence of interaction, the single particle excitation gap is corrected by the interaction, which make $\Delta_0$ become $\Delta$.
  In the following, we would see that  it is the correction that results in the non-zero superfluid density.

\subsection{normal density and superfluid density }

Using Eq.(\ref{272}), the double commutator of x-direction is
\begin{eqnarray}\label{4700}
 &&\langle0|[V_x,[H,V_x]]|0\rangle=\langle0|[V_x,[H_0,V_x]]|0\rangle\notag\\
 &&=\frac{N}{m^{*3}}\langle|0[p_x+p_y\sigma_z+p_x\sigma_x ,i(p^{3}_y+p^{2}_x p_y)\sigma_y]|0\rangle\notag\\
&&=\frac{N}{m^{*3}}\langle0|2(p^{4}_y+p^2_xp^2_y)\sigma_x-2(p_xp^{3}_{y}+p^{3}_{x}p_y)\sigma_z|0\rangle\notag\\
&&=0.
\end{eqnarray}

By Eqs.(\ref{21}), (\ref{41}) and (\ref{4700}), the normal density is
\begin{eqnarray}
&&\frac{\rho_{nxx}}{\rho}=\frac{m\langle0|[V_x,[H,V_x]]|0\rangle}{N\Delta^2}=0.
\end{eqnarray}

Similarly, for y-direction, we have
\begin{eqnarray} \label{49}
&& \langle0|[V_y,[H,V_y]]|0\rangle=\langle0|[V_y,[H_0,V_y]]|0\rangle\notag\\
&&=\frac{N}{m^{*3}}\langle0|[p_y+p_x\sigma_z-p_y\sigma_x ,-i(p^{3}_x+p_x p^{2}_y)\sigma_y]|0\rangle\notag\\
&&=-\frac{N}{m^{*3}}\langle0|2(p^{4}_x+p^2_xp^2_y)\sigma_x+2(p^{3}_x p_{y}+p_{x}p^{3}_y)\sigma_z|0\rangle\notag\\
&&=\frac{2Nk^{4}_{0}}{m^{*3}}=\frac{2N\Delta_{0}^{2}}{m^{*}}
\end{eqnarray}
 By Eqs.(\ref{21}) , (\ref{41}) and (\ref{49}), the normal density is
\begin{eqnarray}\label{s}
&&\frac{\rho_{n,yy}}{\rho}=\frac{m\langle 0|[V_y,[H,V_y]]|0\rangle}{N\Delta^2}=\frac{2m\Delta^{2}_{0}}{m^{*}\Delta^2}\notag\\
&&=\frac{m}{m^*}\frac{2\Delta_0}{G_2+\Delta_0}
\end{eqnarray}

Using Eqs.(\ref{27}) and (\ref{28}), so the superfluid density
\begin{eqnarray}
&&\frac{\rho_{s,xx}}{\rho}=0-0=0,
\end{eqnarray}
and \begin{eqnarray}\label{47}
&&\frac{\rho_{s,yy}}{\rho}=\frac{2m}{m^*}-\frac{m}{m^*}\frac{2\Delta_0}{G_2+\Delta_0}\notag\\
&&=\frac{m}{m^*}\frac{2G_2}{G_2+\Delta_0}.
\end{eqnarray}
Similar calculation shows that $\frac{\rho_{n,xy}}{\rho}=\frac{\rho_{n,yx}}{\rho}=\frac{\rho_{s,xy}}{\rho}=\frac{\rho_{s,yx}}{\rho}=0$.

According to Eqs.(\ref{d}) and (\ref{41}), we can observe that when the interaction is zero, i.e., $G_2=0$, the gap is not corrected, $\Delta=\Delta_0=k^{2}_{0}/m^*$.
Then, by Eqs.(\ref{28}), (\ref{s})  and (\ref{47}), we find that the normal density $\rho_{n,yy}$ exhausts  the whole  f-sum rule of y-direction, i.e., $\frac{\rho_{n,yy}}{\rho}=\frac{2m}{m^*}$. As a result, the superfluid density becomes zero.
On the other hand, in the presence of interaction, the excitation gap is corrected, i.e.,
$\Delta\neq \Delta_0$ and $\frac{\rho_{s,yy}}{\rho}\neq0$, then this correction of the excitation gap by interactions leads to a non-vanishing superfluid density in a flat band BEC.

The normal  density tensor is given by
\begin{align}
\frac{\rho_n}{\rho}\equiv\left(                
 \begin{array}{cccc}   
   \frac{\rho_{n,xx}}{\rho}   &\frac{\rho_{n,xy}}{\rho} \\  
   \frac{\rho_{n,yx}}{\rho}   &\frac{\rho_{n,yy}}{\rho} \\  
\end{array}\right)=\left(                
 \begin{array}{cccc}   
  0   &0 \\  
   0   &\frac{m}{m^*}\frac{2\Delta_0}{\Delta_0+G_2} \\  
\end{array}\right).
\end{align}
The superfluid density tensor is given by
\begin{align}\label{3771}
\frac{\rho_s}{\rho}\equiv\left(                
 \begin{array}{cccc}   
   \frac{\rho_{s,xx}}{\rho}   &\frac{\rho_{s,xy}}{\rho} \\  
   \frac{\rho_{s,yx}}{\rho}   &\frac{\rho_{s,yy}}{\rho} \\  
\end{array}\right)=\left(                
 \begin{array}{cccc}   
   0   &0 \\  
   0   &\frac{m}{m^*}\frac{2G_2}{\Delta_0+G_2} \\  
\end{array}\right).
\end{align}

By Eqs.(\ref{41}) and (\ref{3771}), the superfluid density is related to sound velocity $c(\mathbf{\hat{q}})$, mass density $\rho=mn$ and compressibility $\kappa=\frac{2}{(g+g')n^2}=\frac{1}{nG_1}$:
\begin{eqnarray}\label{kk}
\frac{\rho_{s}(\mathbf{\hat{q}})}{\rho}=\rho c^{2}(\mathbf{\hat{q}})\kappa.
\end{eqnarray}
A similar relation holds in the spin orbit coupled BEC \cite{normaldensity}.
Here using Eq.(\ref{322}), we introduce the superfluid density of $\mathbf{q}$'s direction \cite{yicai2018Josephson, yicai2020},
 \begin{eqnarray}
\rho_{s}(\mathbf{\hat{q}})=\sum_{ij=x,y}\rho_{s,ij}q_{i}q_{j}/q^2=\frac{m}{m^*}\frac{2 G_2sin^2(\theta)}{\Delta_0+G_2}\rho.
\end{eqnarray}

In most cases, the compressibility is greater than zero ($\kappa >0$), indicating thermodynamic stability. Then, by Eq.(\ref{kk}), we can infer that  the non-zero sound velocity is typically associated with a non-vanishing superfluid density. Conversely, when the sound velocity is zero, the superfluid density also becomes zero.

Furthermore, when the interaction energy $G_2=(g-g')n$ is significantly smaller than the excitation gap $\Delta_0$, i.e.,  $G_2/\Delta_{0}\ll 1$, the superfluid density $\rho_{s}(\mathbf{\hat{q}})$ is  directly proportional to the difference between $g$ and $g'$:
 \begin{eqnarray}\label{470}
&&\frac{\rho_{s}(\mathbf{\hat{q}})}{\rho}\simeq\frac{m}{m^*}\frac{2G_2sin^2(\theta)}{\Delta_0}\propto  G_2 \propto g-g'.
\end{eqnarray}

In Ref.\cite{Julku2021B}, the authors only considered the case where  $g\neq0$ and $g'\equiv0$. They found that in a flat band BEC, there exists a non-zero superfluid density.
However, we have discovered that the existence of a non-vanishing superfluid density also depends on the form of the interaction. For example,  in the U(2) invariant case ($g=g'$ and $G_2=0$), then the superfluid density is zero.
This conclusion is based on the assumption that the contribution of quantum fluctuations to the superfluid density is negligible. However, it is worth noting that the contribution of quantum fluctuations is typically much smaller \cite{yanglijun,lianglong,Julku2021B} than the linear terms of the interaction parameters considered in this study.

\subsection{relation to quantum metric }
The  quantum metric  is a  geometric metric tensor of the  Bloch state \cite{Provost}, which is defined by
\begin{align}\label{54}
g_{n,ij}=\langle \frac{\partial u_n}{\partial p_i}|[1-|u_n\rangle \langle u_n |]|\frac{\partial u_n}{\partial p_j}\rangle
\end{align}
where $|u_n\rangle=|u_n(\mathbf{p})\rangle$ is periodic part of Bloch wave function which corresponds the n-th energy band, and $|u_n\rangle \langle u_n |$ is projection operator on a subspace spanned by $|u_n\rangle$.
By Eq.(\ref{277}), the periodic part of two band wave functions are given by
\begin{align}
|u_-(\mathbf{p})\rangle=\left(                 
 \begin{array}{c}   
   \frac{p_x-p_y   }{\sqrt{2(p^{2}_{x}+p^{2}_{y})}} \\  
       \frac{-(p_x+p_y)   }{\sqrt{2(p^{2}_{x}+p^{2}_{y})}}\\    
\end{array}\right),\ \ \ \ |u_+(\mathbf{p})\rangle=\left(                 
 \begin{array}{c}   
   \frac{p_x+p_y   }{\sqrt{2(p^{2}_{x}+p^{2}_{y})}} \\  
       \frac{p_x-p_y   }{\sqrt{2(p^{2}_{x}+p^{2}_{y})}}\\    
\end{array}\right).
\end{align}
The condensation occurs at lower band, by Eq.(\ref{54}), the quantum metric is given by \cite{Iskin2019}
\begin{align}
&g_{-,xx}=\frac{p^{2}_{y}}{(p^{2}_{x}+p^{2}_{y})^2}, \ \ g_{-,yy}=\frac{p^{2}_{x}}{(p^{2}_{x}+p^{2}_{y})^2},\notag\\
&g_{-,xy}=g_{-,yx}=\frac{-p_{x}p_{y}}{(p^{2}_{x}+p^{2}_{y})^2}.
\end{align}

At condensate momentum $\mathbf{p}_0=[p_x,p_y]=[k_0,0]$, the quantum metric is
\begin{align}
&g_{-,xx}=0, \ \ g_{-,yy}=\frac{1}{k^{2}_{0}}=\frac{1}{m^*\Delta_0},\notag\\
&g_{-,xy}=g_{-,yx}=0.
\end{align}

For weakly interacting case, i.e., $G_2/\Delta_0\ll 1$, by Eq.(\ref{470}), we find that the superfluid density is directly proportional to the product of  $G_2$ and quantum metric:
\begin{align}
&\frac{\rho_{s}(\mathbf{\hat{q}})}{\rho}\simeq\frac{m}{m^*}\frac{2G_2sin^2(\theta)}{\Delta_{0}}=2mG_2g_{-}(\mathbf{\hat{q}}),
\end{align}
where we introduce $\mathbf{q}$'s direction quantum metric
\begin{align}
&g_{-}(\mathbf{\hat{q}})\equiv \sum_{ij}g_{-,ij}\hat{q}_i\hat{q}_j=\frac{sin^{2}(\theta)}{m^*\Delta_0}.
\end{align}
Furthermore, one can show that  as a tensor equation, the following relation for superfluid density
\begin{align}
&\frac{\rho_{s,ij}}{\rho}\simeq2mG_2g_{-,ij}.
\end{align}
holds.
When $g'=0$, by Eq.(\ref{41}), the sound velocity is reduced to
\begin{align}
c(\mathbf{\hat{q}})=\frac{gn|sin(\theta)|}{\sqrt{gnm^*+k^{2}_{0}}}.
\end{align}
Furthermore for weakly interacting case, i.e., $gn/\Delta_0\ll 1$,
in terms of quantum metric, we find
\begin{align}
c(\mathbf{\hat{q}})\simeq\frac{gn|sin(\theta)|}{\sqrt{k^{2}_{0}}}=gn\sqrt{g_{-}(\mathbf{\hat{q}})}.
\end{align}
We see that the sound velocity is directly proportional to the product of interaction parameter and the square root of quantum metric, which is consistent with Eq.(15) in Ref. \cite{Julku2021B}.

\subsection{perturbation theory }
 The dependence of superfluid density on interaction parameters suggests that interactions have a perturbative effect on physical quantities.
 In this subsection, we use  the fist-order perturbation theory to calculate the excitation gap $\Delta$. We find that, up to the linear order of interaction parameters,
 this simplest perturbation theory can give the same result for the gap of Eq.(\ref{41}).

 In momentum space, the interaction energy of Eq.(\ref{eqn1}) can be given by
  \begin{align}\label{64}
&V_{int}=\frac{g}{2V}\sum_{\mathbf{k}_1+\mathbf{k}_2=\mathbf{k}_3+\mathbf{k}_4}[c^{\dag}_{1\mathbf{k}_1}c^{\dag}_{1\mathbf{k}_2}c_{1\mathbf{k}_3}c_{1\mathbf{k}_4}+c^{\dag}_{2\mathbf{k}_1}c^{\dag}_{2\mathbf{k}_2}c_{2\mathbf{k}_3}c_{2\mathbf{k}_4}]\notag\\
&+\frac{g'}{2V}\sum_{\mathbf{k}_1+\mathbf{k}_2=\mathbf{k}_3+\mathbf{k}_4}c^{\dag}_{1\mathbf{k}_1}c^{\dag}_{2\mathbf{k}_2}c_{1\mathbf{k}_3}c_{2\mathbf{k}_4}.
\end{align}

We assume that when BEC takes place, all the $N$ particles occupy the lower flat band state described by $\psi_-(\mathbf{p}=k_0 \mathbf{e}_x)$ in Eq.(\ref{277}).
Using second quantization operator, the many-body BEC state can be written as
\begin{align}
|0\rangle=\frac{1}{\sqrt{N!}}(c^{\dag}_{-})^{N}|vac\rangle,
\end{align}
where $|vac\rangle$ is vacuum state and $c^{\dag}_{-}$ is the creation operator for state $\psi_-(\mathbf{p}=k_0 \mathbf{e}_x)$.
When one particle occupies upper band state $\psi_+(\mathbf{p}=k_0 \mathbf{e}_x)$, the remained $N-1$ ones are still in lower band state, then this excited state can be represented by
\begin{align}
|1\rangle=\frac{1}{\sqrt{1!(N-1)!}}c^{\dag}_{+}(c^{\dag}_{-})^{N-1}|vac\rangle,
\end{align}
where $c^{\dag}_{+}$ is the creation operator for state $\psi_+(\mathbf{p}=k_0 \mathbf{e}_x)$.

In the interaction energy Eq.(\ref{64}), taking $\mathbf{k}_1=\mathbf{k}_2=\mathbf{k}_3=\mathbf{k}_4=k_0\mathbf{e}_x$,
using basis transformation [see Eq.(\ref{277})]
\begin{align}\label{2721}
\left(                 
 \begin{array}{c}   
   c_{1k_0\mathbf{e}_x}    \\  
   c_{2 k_0\mathbf{e}_x}      
\end{array}\right) =\frac{1}{2}\left(                 
 \begin{array}{cccc}   
   1  & 1  \\  
   -1 & 1 \\  
\end{array}\right)\left(                 
 \begin{array}{c}   
   c_{-}    \\  
   c_{+}      
\end{array}\right),
\end{align}
representing  the interaction energy with operator $c_{-}$ and $c_{+}$, then we can get
\begin{align}
&V_{int}=\frac{g}{4V}[c^{\dag}_{-}c^{\dag}_{-}c_{-}c_{-}+c^{\dag}_{+}c^{\dag}_{+}c_{-}c_{-}+c^{\dag}_{-}c^{\dag}_{-}c_{+}c_{+}\notag\\
&+c^{\dag}_{+}c^{\dag}_{+}c_{+}c_{+}+4c^{\dag}_{+}c^{\dag}_{-}c_{-}c_{+}]\notag\\
&+\frac{g'}{4V}[c^{\dag}_{-}c^{\dag}_{-}c_{-}c_{-}-c^{\dag}_{+}c^{\dag}_{+}c_{-}c_{-}-c^{\dag}_{-}c^{\dag}_{-}c_{+}c_{+}+c^{\dag}_{+}c^{\dag}_{+}c_{+}c_{+}].
\end{align}

Up to the first order perturbation, the ground state energy and the excited state energy are
\begin{align}
&E_0=\langle 0|H|0\rangle=\langle 0|V_{int}|0\rangle=\frac{(g+g')N(N-1)}{4V},\notag\\
&E_1=\langle 1|H|1\rangle=\frac{k_{0}^{2}}{m^*}+\langle 1|V_{int}|1\rangle\notag\\
&=\Delta_0+\frac{(g+g')(N-1)(N-2)}{4V}+\frac{4g(N-1)}{4}.
\end{align}
In the above equation, we use the fact that the energy of flat band is zero.
Then the excitation gap is given by the difference of two energies, i.e.,
\begin{align}
\Delta\equiv E_1-E_0=\Delta_0+\frac{(g-g')(N-1)}{2V}.
\end{align}
When the particle number $N$ and volume $V$ are very large , $\frac{(N-1)}{V}\simeq \frac{N}{V}= n$, then finally
\begin{align}
\Delta=\Delta_0+\frac{(g-g')n}{2}=\Delta_0+\frac{G_2}{2}.
\end{align}
When interaction energy is much smaller than gap $G_2/\Delta_0\ll1$, the square of gap is given by
\begin{align}
\Delta^2=\Delta_0[\Delta_0+G_2]+O[(G_2/\Delta_0)^2],
\end{align}
which is consistent with gap of Eq.(\ref{41}).
After obtaining the excitation gap using perturbation theory up to the first order of interaction parameters, the normal density and superfluid density can be calculated using the double commutator method, as demonstrated in subsection \textbf{C}.

\section{Conclusion}
In summary, based on the f-sum rule, we propose a double commutator  method to calculate the superfluid density of two band BEC.
We prove that the sum of superfluid and normal density is equal to the weight of the f-sum rule. In addition, the normal density is determined by the average value of a double commutator and the excitation gap.

As an application of this method, we use it to calculate the superfluid density of a 2D flat-band BEC. Using the Bogoliubov method, we calculate the sound velocity and excitation gap, and then the normal and superfluid density. We also present a universal formula that connects the superfluid density, sound velocity, and compressibility, showing that the superfluid density is proportional to the product of the square of the sound velocity and compressibility. Our findings suggest that the existence of non-vanishing superfluid density depends on the form of interaction. For example, in the case of U(2) invariant interaction, the superfluid density vanishes. Additionally, we find that when the interaction is small, the sound velocity and superfluid density are proportional to the interaction parameter and are also related to the quantum metric. In contrast to the quantum metric perspective, the double commutator method shows that the correction of the excitation gap by interactions is the origin of the non-vanishing superfluid density in flat band BEC.

It should be noted that the calculation of the excitation gap using the double commutator method is not limited to the Bogoliubov method. Other methods, such as perturbation theory and the hydrodynamic method, can also be used. It would be meaningful to develop other different methods to study these topics.

\section*{Acknowledgements}
We thank Zhi-Gang Wu and Shizhong Zhang for useful discussions.
This work was supported by the NSFC under Grants No. 11874127, the Joint Fund with Guangzhou
Municipality under No. 202201020137, and the Starting Research Fund from Guangzhou University under
Grant No. RQ 2020083.

\end{document}